# JOINT TIME-SERIES ANALYSIS OF HOMESTAKE AND GALLEX-GNO SOLAR NEUTRINO DATA: EVIDENCE FOR RIEGER-TYPE PERIODICITIES AND THEIR INTERPRETATION AS R-MODE OSCILLATIONS

P.A. Sturrock[1]

[1]Center for Space Science and Astrophysics, Stanford University, Stanford, CA 94305-4060

## ABSTRACT

Previous power-spectrum analyses of the Homestake and GALLEX-GNO radiochemical solar neutrino data have shown evidence of the effects of solar rotation and also indications of oscillations with two periods that are well known in solar physics: one at about 154 days period (the Rieger periodicity) and another at about 52 days period. In this article, we make a systematic investigation of both data sets, taken together, in a manner designed to test the hypothesis that the above periodicities, and a related periodicity with a period of about 77 days, are due to r-mode oscillations with spherical harmonic indices $l = 3$ and $m = 1, 2$, and $3$. These would have frequencies of $\nu_R/6$, $\nu_R/3$, and $\nu_R/2$, where $\nu_R$ is the sidereal rotation rate where the waves are excited. We first form a joint statistic (of second order) that combines information from power spectra formed from Homestake and GALLEX-GNO data. From this statistic, we then form another joint statistic (of fourth order) designed to search frequency space for evidence of a sidereal rotation frequency that leads simultaneously to four peaks in the Homestake-GALLEX-GNO spectrum: one corresponding to the synodic rotation frequency (at $\nu_R - 1 \text{ y}^{-1}$), and the other three corresponding to the $l = 3$ r-modes. We find that, in the range $0 - 20 \text{ y}^{-1}$, there is one notable peak at $13.88 \text{ y}^{-1}$ which falls in the rotation frequency range of the radiative zone. By applying the shuffle test to the data, we find that the probability of obtaining by chance a peak this big or bigger in the rotation-frequency band of the radiative-zone is less than 0.1%. This procedure identifies the following four peaks in the Homestake-GALLEX-GNO joint spectrum: one at $12.88 \text{ y}^{-1}$, which we identify with the synodic rotation frequency (period 28.4 days, which has been previously recognized in the Homestake data), and others at 2.33, 4.62, and 6.94 $y^{-1}$ (periods of 157, 78, and 52 days), which correspond to the $l = 3$, $m = 1$, 2, and 3 r-mode frequencies. By comparing this rotation frequency with helioseismology data, we infer that the r-modes are occurring just below the tachocline. We speculate that the excitation of these r-modes is due to their symbiotic interaction with retrograde Alfven waves propagating in toroidal magnetic flux tubes which may be related to the solar dynamo. We argue that this combination of waves can be unstable, leading to the growth of both waves by a process analogous to the instability that occurs in neutron stars due to the coupling of r-modes to gravitational waves. The modulation of the neutrino flux is attributed to the poloidal motion of a magnetic structure that moves in and out of the narrow beam of neutrinos that originate in the solar core and are detected on Earth.

INTRODUCTION

In recent articles, we have presented the results of our analyses of Homestake (Davis & Cox 1991; Lande et al. 1992; Cleveland et al. 1995, 1998) and GALLEX-GNO (Anselmann et al. 1993, 1995; Hampel et al. 1996, 1997; Altman et al. 2000) solar neutrino data. The main focus in these articles has been the search for evidence of modulation that may be attributed to solar rotation. In GALLEX-GNO data, we find a modulation at 13.59 $y^{-1}$ (period 26.88 days), corresponding to a sidereal rotation rate of 14.59 $y^{-1}$ (Sturrock & Weber 2002), which matches the rotation rate as determined by helioseismology (Schou et al. 1998) at or near the equator in the convection zone at a normalized radius of about 0.85. In Homestake data, we find the same modulation but we also find a stronger modulation at 12.88 $y^{-1}$ (period 29.1 days), corresponding to a sidereal rotation frequency of 13.88 $y^{-1}$ (Sturrock, Walther, & Wheatland 1997), compatible with estimates of the rotation rate of the radiative zone made by helioseismology. In addition to these frequencies that may be related to the Sun's internal rotation, we have noticed some other interesting periodicities in solar neutrino data.

The tendency for gamma-ray flares to repeat with a period of 154 days (frequency about 2.4 $y^{-1}$) was discovered by Rieger et al. (1984). Investigators have reported that this periodicity is evident in general flare data and in daily sunspot area measurements (see, for instance, Bai & Cliver 1990 and references therein) and that there is evidence also for additional similar periodicities, notably one with period about 77 days (about 4.7 $y^{-1}$; Bai 1992) and one with period about 51 days (about 7.1 $y^{-1}$; Bai 1994; Kile & Cliver 1991). Some of the suggested interpretations of these periodicities are to be found in a recent article by Ballester, Oliver, & Carbonell (2002).

We have previously noted that a periodicity of about 2.3 $y^{-1}$ has been found in Homestake data (Sturrock, Walther, & Wheatland 1997). We have also noted that a periodicity of about 6.9 $y^{-1}$ has been found in GALLEX data (Sturrock et al. 1999), and suggested that Rieger-type oscillations may be due to internal r-mode oscillations (Papaloizou & Pringle 1978; Provost, Berthomieu, & Rocca 1981; Saio 1982) which are retrograde waves similar to Rossby waves in the Earth's atmosphere (Pedlosky1987). Lou (2000) and Wolff (1992) have recently presented evidence that r-modes play a role in solar dynamics. In this article, we explore the hypothesis that Rieger-type periodicities are due to r-mode oscillations that modulate solar activity and the solar neutrino flux, and we speculate on the excitation mechanism of these oscillations.

In terms of the usual spherical harmonic indices ($l$, $m$, and $n$), r-mode frequencies, as they would be observed in the Sun itself, are given by

$$\nu(l,m,Sun) = \frac{2m\nu_R}{l(l+1)} \tag{1.1}$$

where $l \geq 2$, $m = 1,...,l$, and $\nu_R$ is the absolute (sidereal) rotation frequency. It is significant that the frequencies do not depend upon the radial index $n$. To an observer in space, the frequencies would be reduced by $m\nu_R$. To an observer on Earth, the frequencies would be given by

$$\nu(l,m,Earth) = \frac{2m\nu_R}{l(l+1)} - m(\nu_R - 1) . \tag{1.2}$$

If an observer on Earth detects the frequencies given by (1.1), rather than the frequencies given by (1.2), it must be because the r-modes are interacting with some source, such as a magnetic structure, that rotates with the Sun. We return to this point later in this article.

The error bars on radio-chemical measurements of the solar neutrino flux are quite large. Hence we face the challenge of extracting a presumably small signal from assuredly high-amplitude noise. It is well known that for a given noise level and given signal amplitude, the signal-to-noise ratio may be improved by examining a more extensive time series. Unfortunately, the amount of neutrino data is limited.

However, there are two other ways by which we may improve the signal-to-noise ratio. (a) We may carry out a combined analysis of both Homestake and GALLEX-GNO data sets. (b) We may search for modulation at a rotation frequency and also at a related set of r-mode frequencies. Since the Rieger-type periodicities correspond to *l = 3, m = 1,2,3*, this amounts to a search for four related periodicities in two independent data sets. In this way, we should obtain an improvement in the signal-to-noise ratio that is comparable to (but not equal to, since some phase information is missing) the improvement that would be obtained by analyzing a single periodicity in four sequences of 108 runs extending over 24 years (Homestake), and in four sequences of 84 runs extending over 9 years (GALLEX-GNO), i.e. to the analysis of a single periodicity in a data set of 768 runs extending over 132 years.

In Section 2, we outline a recently developed procedure (Sturrock 2003) for the joint analysis of two or more power spectra. We use this technique to combine the Homestake and GALLEX-GNO spectra in Section 3, and use it again to search for evidence of r-modes in Section 4. We briefly explore a possible excitation mechanism of r-mode oscillations in Section 5, and offer additional discussion in Section 6.

## 2. JOINT SPECTRUM STATISTIC

We briefly review a recently proposal for the joint analysis of power spectra derived from two or more time series (Sturrock 2003). If a time series is properly normalized, and if the null hypothesis is that the data are derived from normally distributed random noise, the probability distribution function (pdf) for the power S is given by

$$P(S)dS = e^{-S}dS, \qquad (2.1)$$

and the cumulative distribution function (cdf), that gives the probability of obtaining a value S or more on the null hypothesis, is given by

$$C(S) = \int_S^\infty e^{-s}ds = e^{-S}. \qquad (2.2)$$

(See, for instance, Scargle 1982.)

Suppose that, from given spectra $S_1(\nu)$ and $S_2(\nu)$, we form the geometric mean

$$X = (S_1 S_2)^{1/2}. \qquad (2.3)$$

Then X is in some sense a measure of the correlation between the two spectra. However, we need a procedure for assessing the significance of this quantity. To facilitate this, we find a function of X that is distributed in the same way as the power of a single time series.

We find that the cumulative distribution function for X is given by

$$C(X) = \int_Z^\infty dz\, P(z) = \int_0^\infty dx\, e^{-x - X^2/x}.\tag{2.4}$$

Hence, if we form

$$J(X) = -\ln(C(X)),\tag{2.5}$$

J will have the same form as the cdf of S. This "joint spectrum statistic" is given explicitly by

$$J(X) = \ln(2X K_1(2X)),\tag{2.6}$$

where $K_1$ is a modified Bessel function.

We may form similar statistics of higher order by replacing $(S_1 S_2)^{1/2}$ in (2.3) by the third or fourth root of the product of three or more power measurements, respectively. There are no simple analytical expressions for the higher-order statistics, but joint spectrum statistics of the third and fourth orders have been computed and plotted (Sturrock 2003). They are shown, together with the second-order statistic, in Figure 1.

## 3. JOINT ANALYSIS OF THE HOMESTAKE AND GALLEX-GNO DATA

We follow the procedure outlined in Sturrock and Weber (2002) in forming power spectra from Homestake and GALLEX-GNO data. We first normalize the flux measurements $g_r$ as follows:

$$x_r = \frac{g_r - mean(g)}{std(g)}, \qquad (3.1)$$

and apply the following operation, designed to minimize the effect of "outliers,"

$$y_r = \tanh(x_r). \qquad (3.2)$$

We denote by $t_r$ the time of each data point (here chosen to be the end time of each run), normalize $y_r$ to have mean value zero, compute the standard deviation

$$\sigma = std(y), \qquad (3.3)$$

and then carry out the Lomb-Scargle power-spectrum analysis (Lomb 1976; Scargle 1982) to arrive at the power spectra $SH(\nu)$, $SG(\nu)$, for Homestake and Gallex-GNO data, respectively. The spectra formed in this way are very similar to those presented in earlier publications (Sturrock, Walther and Wheatland 1997; Sturrock and Weber 2002).

## 4. SEARCH FOR R-MODE OSCILLATIONS

We now apply the operation of equation (2.6) to the product of these two power spectra. The result is shown in Figure 2. The most significant peak in the range 0-20 $y^{-1}$ is to be found at 6.08 $y^{-1}$. This is not obviously due to a systematic effect (although the frequency corresponds to a period close to 60 days), and is therefore an enigma. The next most significant peak is at 13.59 $y^{-1}$, which we discussed in Sturrock and Weber (2002) and attribute to modulation occurring in the convection zone. The four peaks indicated by arrows will be discussed in Section 4.

We wish to consider the possibility that r-modes occur in some region inside the Sun, with corresponding modulation of the neutrino flux. We consider, in particular, the three modes $l = 3$, $m = 1, 2, 3$, since these offer a potential explanation of the three principal Rieger-type periodicities. As we see from equation (1.1), these have frequencies $v_R/6$, $v_R/3$, and $v_R/2$. We wish to consider also the possibility that some structure in this region modulates the neutrino flux. This would be seen as a peak at the synodic frequency $v_R - 1$. We therefore form the product

$$X = \left[ J_2(v_R/6) J_2(v_R/3) J_2(v_R/3) J_2(v_R - 1) \right]^{1/4} . \tag{4.1}$$

We denote by $J_4(v)$ the fourth-order joint spectrum statistic formed from X, and evaluate this using Figure 1. This statistic is displayed, as a function of frequency, in Figure 3. We see that there is one notable peak, for which J = 9.7, at the frequency 13.88 $y^{-1}$. The corresponding synodic frequency, 12.88$y^{-1}$, is the frequency first identified in our analysis of Homestake data (Sturrock, Walther, and Wheatland 1997), and is compatible with the rotation rate of the solar radiative zone (Schou, et al. 1988). For the sidereal frequency 13.88 $y^{-1}$, the four frequencies that play a role in (4.1) are identified by arrows in Figure 2.

We need to assess the possibility that the peak in Figure 3 may have occurred by chance. The rotation frequency of the radiative zone, as determined by helioseismology, (Schou, et al, 1987), falls in the band 13.5 to 14 $y^{-1}$. We carried out 10,000 simulations of the operation described in section 3 and 4, shuffling the Homestake and Gallex-GNO data by randomly reassigning flux measurements to different runs. We searched the wider band 13 – 14 $y^{-1}$, and found only 16 simulations for which the band contains a peak as big as or bigger than that shown in Figure 3. This indicates that less than 1 in 1,000 simulations would have a comparable or larger peak in the radiative-zone rotation frequency band. We conclude that the apparent association of the peak in Figure 3 with the radiative zone may have occurred by chance with probability less than 0.1%. The peak therefore appears to be real.

We have analyzed Homestake and Gallex-GNO separately by the above procedure, replacing $J_2(\nu)$ in (4.1) by $SH(\nu)$ and $SG(\nu)$ in turn. From Homestake data, we find a peak of 6.7 at 13.88 y$^{-1}$, with a 3 db half-width of 0.02y$^{-1}$. From Gallex-GNO data, we find a peak of 6.5 at 13.81y$^{-1}$, with a 3 db half-width of 0.04y$^{-1}$. Hence the peak in Figure 3 occurs independently, but with lower significance, in both Homestake and Gallex-GNO data taken separately. For $\nu_R = 13.88 \, y^{-1}$, the three r-mode frequencies are 2.31, 4.63, and 6.94 y$^{-1}$, and the corresponding synodic frequency is 12.88 y$^{-1}$. (These are the four frequencies that are indicated by arrows in Figure 1.) We see that a peak occurs at each of these frequencies in the joint spectrum statistic formed from the Homestake and Gallex-GNO power spectra.

## 5. POSSIBLE EXCITATION MECHANISM OF R-MODE OSCILLATIONS

We now speculate on the possible excitation mechanism of r-mode oscillations inside the Sun. For a rigidly rotating star, r-modes are known to be stable (see, for instance, Saio 1982). It is conceivable that r-modes might be unstable in a star with a more complex rotation profile, such as the Sun. However, one would then expect the r-mode excitation to be steady. Assuming that the Rieger-type oscillations are indeed due to r-modes, we know that the excitation of these modes is not steady. Another important point is that the Rieger-type oscillations have an important influence on the Sun's internal magnetic field, as evidenced by the fact that these oscillations show up in various indices of solar activity.

One may get a hint of a possible resolution of this problem by recalling that r-modes are known to be unstable in neutron stars (Andersson 1998; Andersson, Kokkotas, & Sterliouglas 1999; Friedman & Morsink 1998; Lindblom, Owen, & Morsink 1998). In that situation, the instability results from the coupling of m = 2 r-mode oscillations to gravity waves. It is hardly conceivable that gravity waves will play any role in such a large and slowly rotating star as the Sun. However, there may be some other wave that interacts symbiotically with r-modes in the Sun, in such a way that the combination is unstable and both waves grow to large amplitudes.

If the second wave is magnetohydrodynamic, we are likely to find plausible answers to the two points raised above. The Sun's internal magnetic field varies with the solar cycle, and from cycle to cycle, so it would be quite reasonable that the wave-wave interaction would behave in a solar-cycle-dependent manner. Furthermore, if the Sun's magnetic field is interacting with r-modes, it is not unreasonable that the result of this interaction might be manifested at the photosphere.

Hence we need to consider two questions:
(a) What type of MHD wave might interact with r-modes, and
(b) Is there reason to believe that the interaction would be unstable, and lead to the growth of both waves?

Since r-modes have patterns that rotate about the Sun's rotational axis, we need to look for MHD waves that might do the same. Following Babcock (1961) and Leighton (1964, 1969), most dynamo theories involve the possibility that a toroidal magnetic flux system is located somewhere near the tachocline (see, for instance, Krause, Raedler, & Ruediger 1993). Hence it is reasonable to consider that r-modes may be interacting with Alfven waves that propagate around the Sun.

Let us consider, to be specific, the l = 3, m = 1, r-mode, which appears to be responsible for the Rieger oscillation with a period of about 154 days. We see from (1.1) that the r-mode frequency, in the frame of the rotating Sun, will be $v_R/6$. This is a retrograde wave. A retrograde Alfven wave will be in synchronism with the r-mode wave if the Alfven speed is given by

$$v_A = 2\pi (v_R/6) r , \qquad (5.1)$$

where r is the radius of the region where the waves are excited. At the bottom of the convection zone, r = $10^{10.69}$ cm, and $v_R = 14.2 \, y^{-1} = 10^{-6.35} \, s^{-1}$, so that the required value of the Alfven speed is $10^{4.36}$ cm s$^{-1}$. The Alfven speed is related to the magnetic field strength B and the density $\rho$ by

$$v_A = \frac{B}{(4\pi\rho)^{1/2}} . \qquad (5.2)$$

Since $\rho \approx 0.08\, g\, cm^{-3}$ at the bottom of the convection zone, we find that the retrograde Alfven wave will be in synchronism with the r-mode wave if the magnetic field strength is about 23,000 gauss. If the m = 2 and m = 3 modes occur in the same region, they would require field strengths of about 46,000 gauss and 70,000 gauss, respectively. These values are in the range of field strengths that one expects to find in that region. It is possible that only one flux tube will be active at any one time, but it is also possible that two different flux tubes, with different field strengths, might be active at the same time. Of course, it is not necessary that the flux tube be strictly symmetric about the solar axis. In fact, as we noted in Section 1, the fact that solar time-series show periodicities at the frequencies given by (1.1) already indicates that the r-modes are interacting with a structure, presumably magnetic, that rotates with the Sun. An Alfven wave could propagate around an inhomogeneous flux tube and still determine a well-defined mode propagating around the Sun. If there are several flux tubes in the same region, an r-mode would interact most strongly with the tube that has a phase velocity closest to that of the r-mode.

An important remaining question is whether the interaction of these two waves is likely to be unstable, leading to the growth of both waves. We believe that it will be, for the following reason. It has been known for some time that, in linear wave theory, a wave can act as if it has negative energy (Sturrock 1960) and that some plasma instabilities can be understood on this basis (Sturrock 1962): if a wave is known to have positive energy in one frame, it will appear to have negative energy if it is viewed from a frame moving with such velocity that the phase velocity of the wave is reversed in sign. Suppose that the Alfven speed is slightly smaller than the speed of the r-mode, and consider a reference frame rotating with a speed intermediate between the speeds of the two waves. In this frame, the r-mode still moves in a retrograde sense (albeit with a small speed), so it will still have positive energy. However, the retrograde Alfven wave will have a small velocity in the prograde sense, so it will appear to have negative energy. Hence we should expect the interaction of these two waves to be unstable, leading to the growth of both waves. This is similar to the mechanism of two-stream instabilities in plasmas (Sturrock 1962, 1994). It is also similar to the mechanism that leads to the instability of r-modes interacting with gravity waves in neutron stars, in which gravity waves have positive energy in the rest frame but r-modes have negative energy since they have a phase velocity in a fixed frame that is opposite to their phase velocity in the frame of the rotating star.

## 6. DISCUSSION

If the feature found in Figure 3 represents a real solar phenomenon, it seems likely that something similar would show up in a similar analysis of other solar data. We have therefore carried out a similar analysis of sunspot data for the time period 1970-1999, inclusive, corresponding to the interval for which radiochemical neutrino data are available. We have formed a power spectrum of the sunspot data by the Rayleigh-power procedure (Mardia 1972; Droge et al. 1990), and repeated the analysis of section 4. The result is shown in Figure 4, which shows a very strong feature in the neighborhood of 14 $y^{-1}$. However, this may be due in large measure to the strong peaks in the sunspot power spectrum in the neighborhood of 13 $y^{-1}$ due to solar rotation. We have therefore repeated the analysis, omitting the synodic term in equation (4.1), retaining only the three r-mode terms. The result is shown in Figure 5. We see that there is still a significant structure in the neighborhood of 14 $y^{-1}$. However, the structure in Figures 4 and 5, derived from sunspots, is more ragged than the structure in Figure 3, derived from neutrino data. This is consistent with the picture that Rieger-type oscillations are caused by r-mode oscillations deep in the convection zone or near the tachocline, where neutrino modulation may take place. It is also interesting to note that this picture helps us understand why l = 3 modes are more obvious than l = 2 or l = 4 modes: modes with odd l-values have a nonzero poloidal velocity at the equator that could move magnetic regions into and out of the path of neutrinos traveling from the solar core to Earth.

The set of r-modes found in our analysis of neutrino data corresponds to a sidereal rotation frequency of 13.88 $y^{-1}$, which is the rotation frequency at a normalized radius of about 0.68, just below the tachocline (Schou et al. 1998). The corresponding r-mode periods, 158, 79, and 53 days, are close to the Rieger-type periods that show up in the time-variation of indices of solar activity. However, the periods are often closer to 154, 77, and 51 days, corresponding to a "parent" frequency of 14.23 $y^{-1}$, which is the sidereal rotation frequency at a normalized radius of about 0.71, just above the tachocline. Hence the range of variation of the Rieger-type periodicities may be understood on the assumption that the r-modes are excited close to the tachocline, but not always at exactly the same radius.

Although we have concentrated in this article on r-mode frequencies as they are manifested in the frame of the rotating Sun given by equation (1.1), due to their association with the Rieger-type frequencies, there is some evidence that the retrograde frequencies given by equation (1.2) also show up in neutrino data. For the sidereal rotation rate 13.88 $y^{-1}$ identified in Section 4, the retrograde frequencies for the l = 3, m = 1, 2, and 3 modes are 10.57, 21.13, and 31.70 $y^{-1}$. In examining the statistic J2 formed from Homestake and GALLEX-GNO power spectra in Section 3, we find peaks at 10.53, 21.17, and 31.55 $y^{-1}$. This is suggestive evidence for r-mode retrograde waves. This point will be examined more systematically in a later publication.

It appears from this analysis that there is evidence that the principal Rieger-type periodicities show up in solar neutrino data, and that these oscillations may be due to r-mode oscillations. If the solar neutrino flux is indeed variable on the time scale of the solar-rotation period, this variability is most probably attributable to the influence of the Sun's internal magnetic field on neutrino propagation. This may be due to the RSFP (resonant spin-flavor precession) effect proposed by Akhmedov (1988a,1988b) and by Lim and Marciano (1988), possibly acting together with the MSW effect (Mikhevev & Smirnov 1986a, 1986b, 1986c; Wolfenstein 1978, 1979), or by some other process that has not yet been proposed.

This research was supported by NSF grant ATM-0097128.

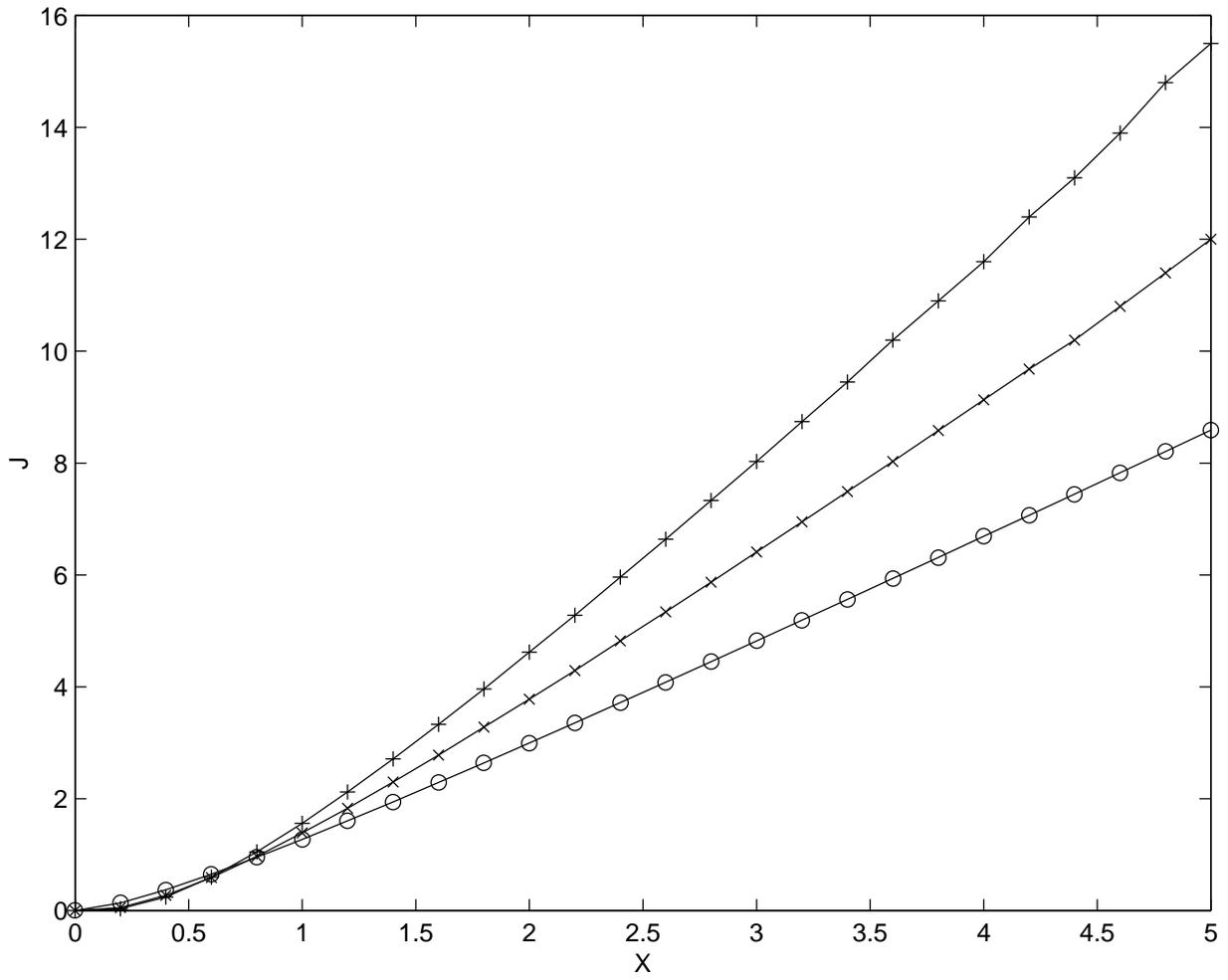

Fig. 1. Joint spectrum statistics of orders 2 ('o'), 3 ('x'), and 4 ('+').

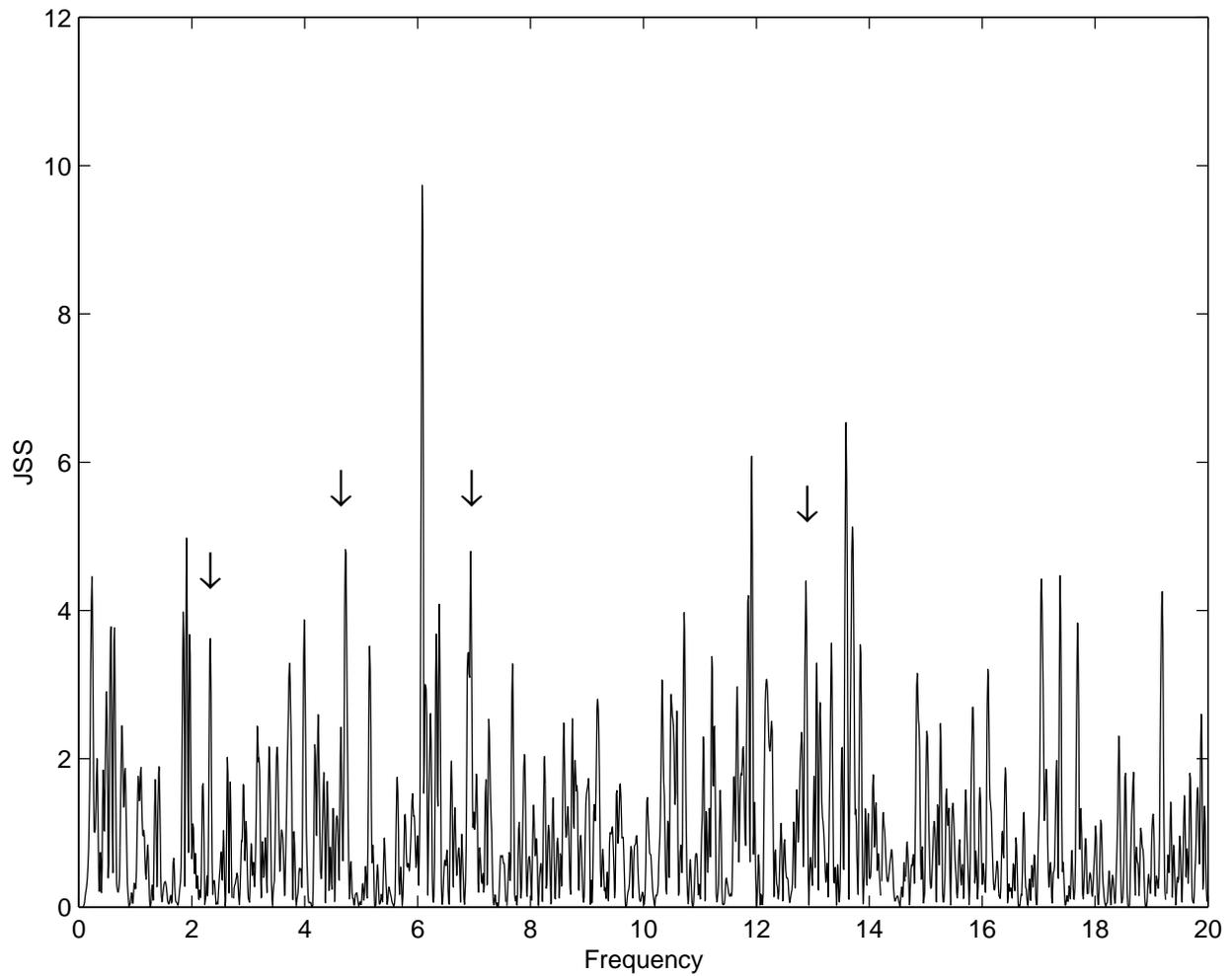

Fig. 2. Joint spectrum statistic of second order formed from power spectra formed from Homestake and GALLEX-GNO data. The arrows indicate the r-mode frequencies and the synodic rotation frequency inferred from the analysis of Section 3.

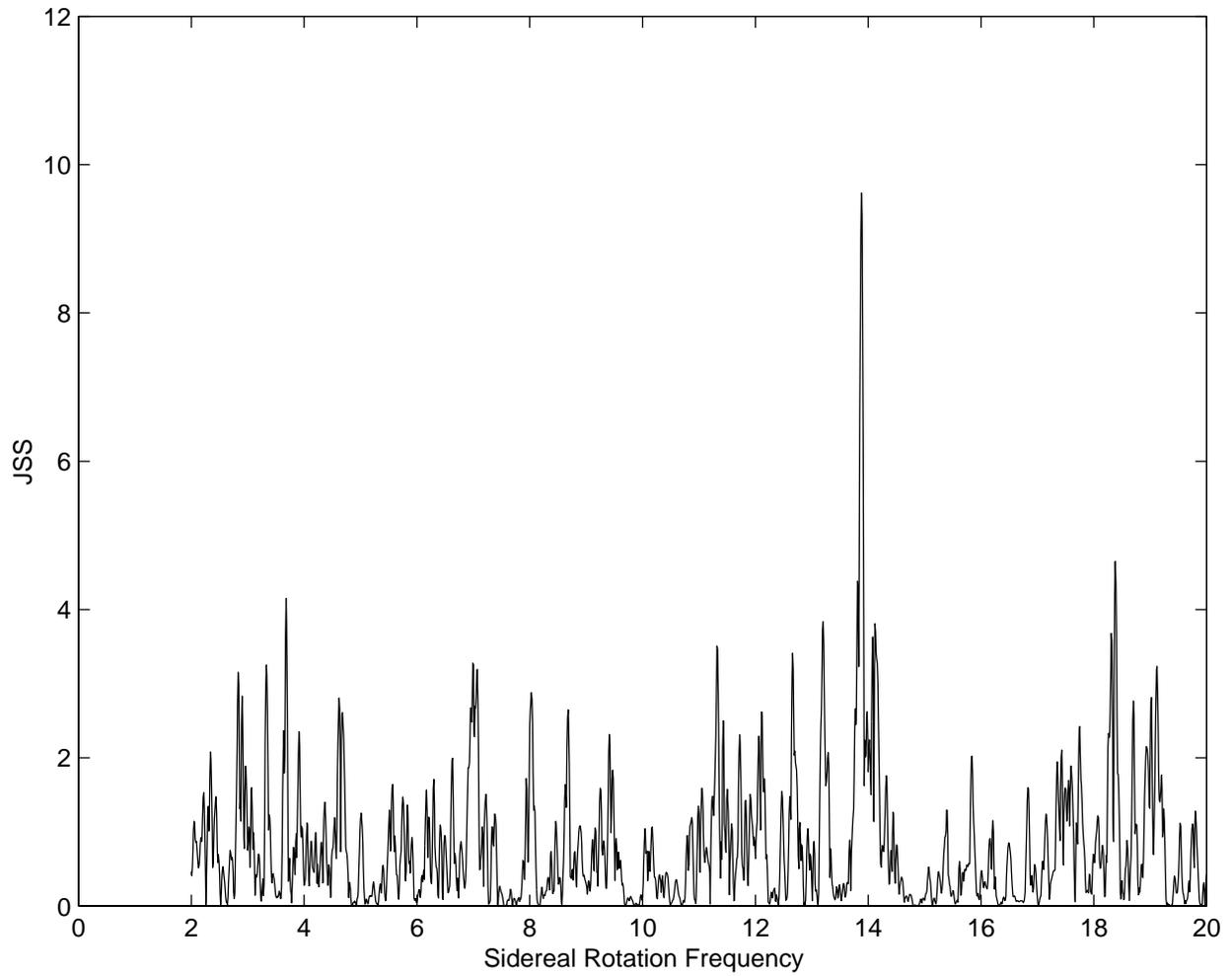

Fig. 3. Joint spectrum statistic of fourth order formed from the Homestake-GALLEX-GNO joint spectrum statistic evaluated at three r-mode frequencies and the synodic rotation frequency.

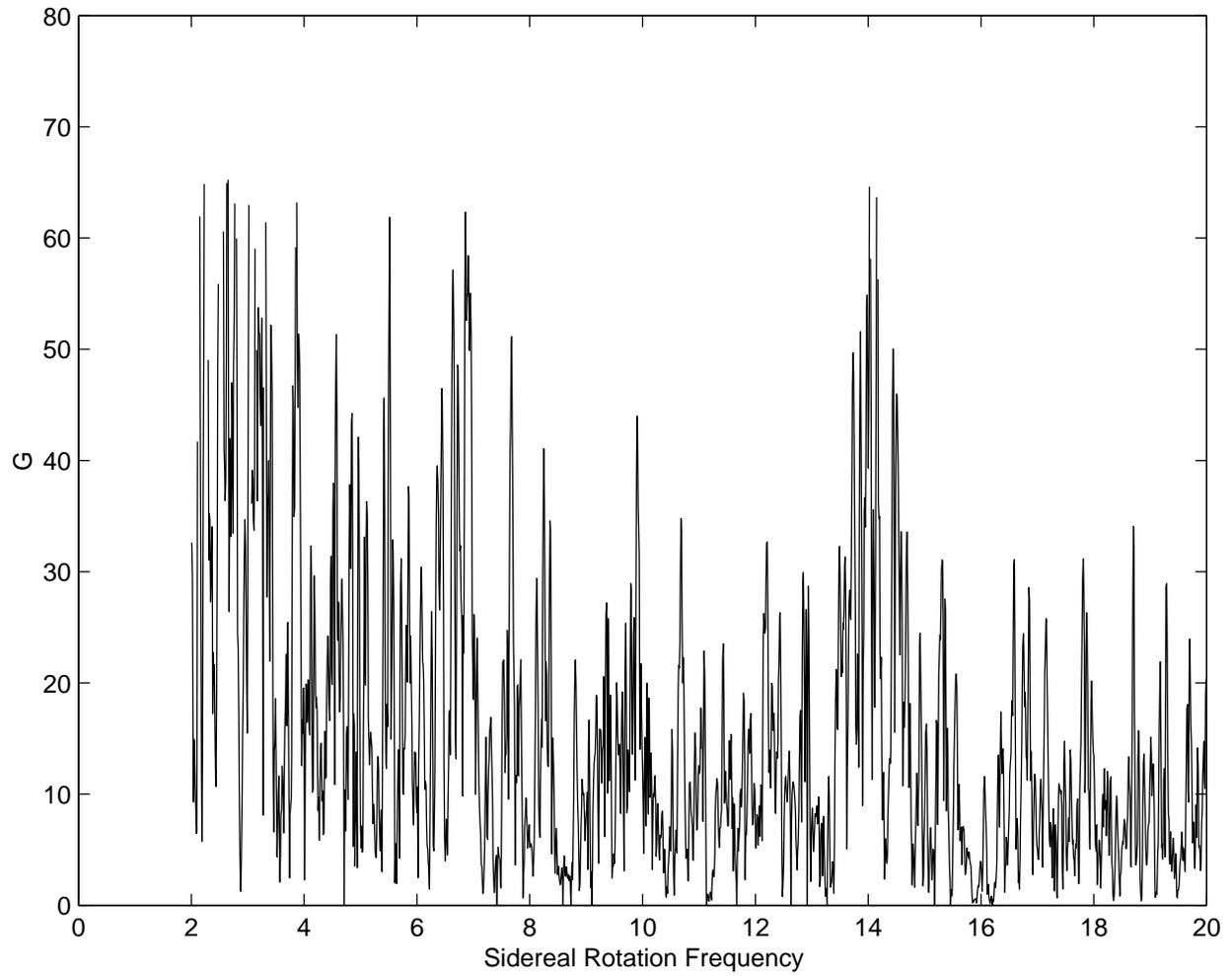

Fig. 4. Joint spectrum statistic of fourth order formed from the Rayleigh power spectrum of the sunspot number for the years 1970 – 1999, inclusive.

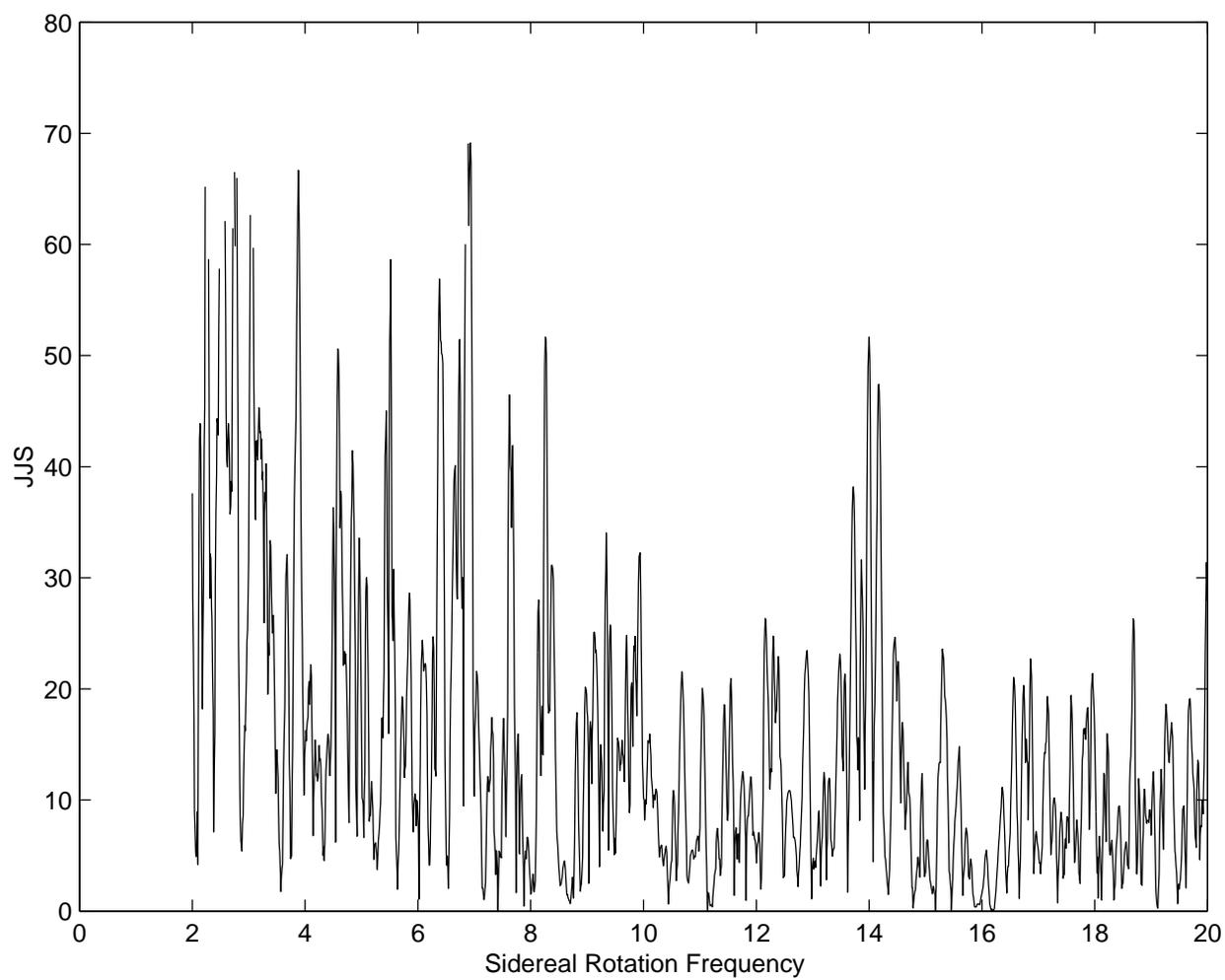

Fig. 5. Joint spectrum statistic of third order formed from the Rayleigh power spectrum of the sunspot number for the years 1970 – 1999, inclusive.